\input{aipcheck}

\documentclass[
    ,final            
  ]
  {aipproc}

\layoutstyle{6x9}

\def \src {IGR~J11215--5952}

\def \hcm {\hbox {\ifmmode $ atom cm$^{-2}\else atom cm$^{-2}$\fi}}

\begin{document}

\title{INTEGRAL observations of IGR~J11215--5952: the first Supergiant Fast X--ray Transient displaying periodic outbursts}



\author{L. Sidoli}{
}

\author{A. Paizis}{
}

\author{S. Mereghetti}{
  address={INAF, Istituto di Astrofisica Spaziale e Fisica Cosmica, Milano, Italy} 
}

\begin{abstract}
The hard X--ray source \src, discovered with INTEGRAL
during a brief outburst in 2005, has been proposed as a new member of
the class of Supergiant Fast X-ray Transients.
Analysing archival INTEGRAL observations of the source field, 
we have discovered two previously unnoticed 
outbursts (in July 2003 and in May 2004), 
spaced by intervals of $\sim$330 days, suggesting a possible orbital period.
The 5--100~keV spectrum is well described by a cut-off
power law, with a photon index of $\sim$0.5, and a cut-off energy
$\sim$15--20~keV, typical of High Mass X--ray Binaries containing a
neutron star. The luminosity is
$\sim$3$\times$10$^{36}$~erg~s$^{-1}$ assuming
6.2~kpc, the distance of the likely optical counterpart, the blue supergiant  HD~306414.
A fourth outburst was discovered in 2006 with XTE/PCA, 329~days
after the third one, confirming the periodic nature of the source 
outbursts.
Follow-up observations with Swift/XRT refined the source position
and confirmed the association with HD~306414. 
The 5--100~keV spectrum, the recurrent nature
of the outbursts, the  blue
supergiant companion star HD 306414, support the hypothesis  that \src\ is a 
Supergiant Fast X--ray Transient, and it is the first object of
this class of High Mass X--ray Binaries displaying periodic outbursts. 
\end{abstract} 

\maketitle


\section{Introduction}

The INTEGRAL satellite has discovered more than
one hundred hard X--ray (E$>15$~keV) sources since its launch in 2002
(see e.g. the 2nd IBIS catalog, \cite{Bird2006}). 
A large fraction of these objects (about 30\%) have been identified
as  High Mass X--ray Binaries (HMXRBs), either
thanks to their association with  blue supergiants or Be stars, 
or based on their X--ray properties typical of HMXRBs, like e.g. periodic pulsations
or hard X--ray spectra (photon index around 
0.5--1 in the 2--10 keV energy range).

Interestingly, INTEGRAL  has discovered a growing  number
of members of two different classes of HMXRBs: highly absorbed persistent sources (with
properties similar to Vela X--1), 
and transient sources displaying brief outbursts, with
a duration of few hours, 
significantly shorter than the transient Be/XRBs outbursts (the Supergiant Fast X--ray
Transients, SFXTs; e.g. \cite{Negueruela2005a},  \cite{Sguera2005})

\src\ is a transient hard X--ray source discovered with the
INTEGRAL satellite in  April 2005 \cite{Lubinski2005}
and tentatively associated with the supergiant star HD~306414
\cite{Negueruela2005b}. 
Here we report on the discovery of two previous unnoticed outbursts
from this source through the analysis of archival INTEGRAL observations
(for details see \cite{SidoliPM2006}).

\begin{figure}[ht!]
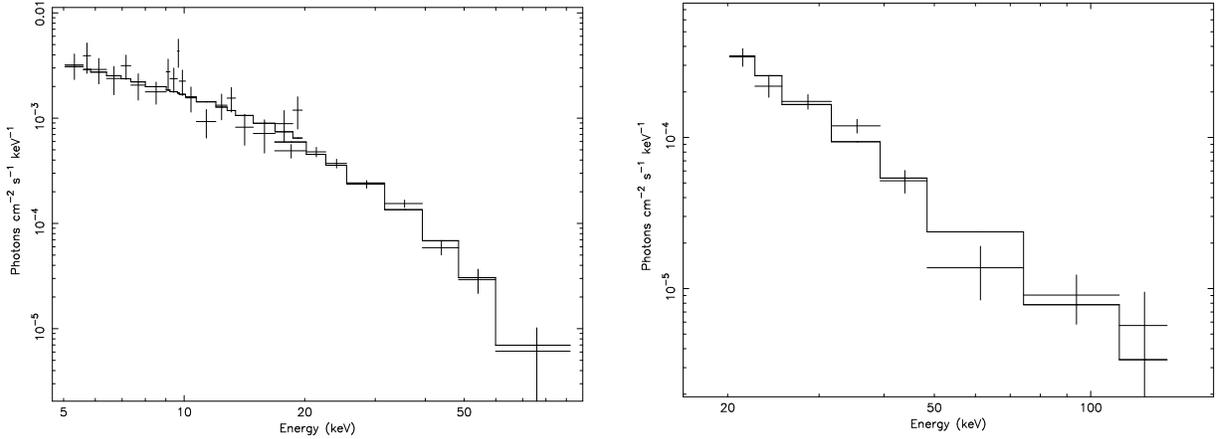

\hbox{\hspace{.3cm}
\includegraphics[height=.35\textheight,angle=-90]{lsidoli_1_fig1a.ps}
\hspace{.5cm}
\includegraphics[height=.35\textheight,angle=-90]{lsidoli_1_fig1b.ps}}
\caption{{\em Left:}  \src\  photon spectrum  from the 2003
outburst (JEM-X and ISGRI, 5--100 keV) fit with a cutoff power
law. {\em Right:}  \src\  photon spectrum  (only ISGRI data,
20--100~keV) from the 2004 outburst, deconvolved with a single power law
(see text). }
\end{figure}

\section{Observations and Results}

The ESA INTEGRAL gamma-ray observatory was launched in October 2002
and
carries three co-aligned coded mask telescopes: the imager IBIS
(\cite{Ubertini2003}), which allows high angular resolution
imaging over a large field of view (29$^{\circ}\times29^{\circ}$)
in the energy range 15\,keV--10\,MeV, the spectrometer SPI
(\cite{Vedrenne2003}; 20 keV--8\,MeV) and the X-ray monitor JEM-X
(\cite{Lund2003}; 3--35\,keV).  

The sky region of \src\ was repeatedly observed by INTEGRAL. We
analyzed all the public IBIS/ISGRI observations pointed within
15$^{\circ}$ of the source.
This translates into  850 individual pointings 
performed between December 2002 and August 2004,
yielding  a total exposure time of about 1.8~Ms. 

\src\ was  detected  in
the 17--40 keV range in  17 pointings, which
correspond to two outbursts occurring on 3-4 July 2003 and
26-27 May 2004. 
The first of these two outbursts was independently discovered  
also by Sguera et al. (\cite{Sguera2006}).
The sparse sample allows us only to put lower limits on the outbursts durations,
which are $\sim$9 hours and about two days for the 2003
and 2004 outbursts, respectively. Both outbursts did not exceed $\sim$7~days.

Our refined estimate for the source position has been obtained
summing together the pointings from both outbursts:
R.A. (J2000)= 11$^h$~21$^m$~50.8$^s$,
Dec.= --59$^{\circ}$~52$'$~48.3$''$, with a statistical uncertainty of 1.2$'$.
This position is consistent with that obtained during the April
2005 discovery outburst (\cite{Lubinski2005}) and the refined uncertainty
region still includes the proposed optical counterpart 
HD~306414 (\cite{Negueruela2005b}, \cite{Masetti2006}).

\begin{figure}
  \includegraphics[height=.70\textheight,angle=-90]{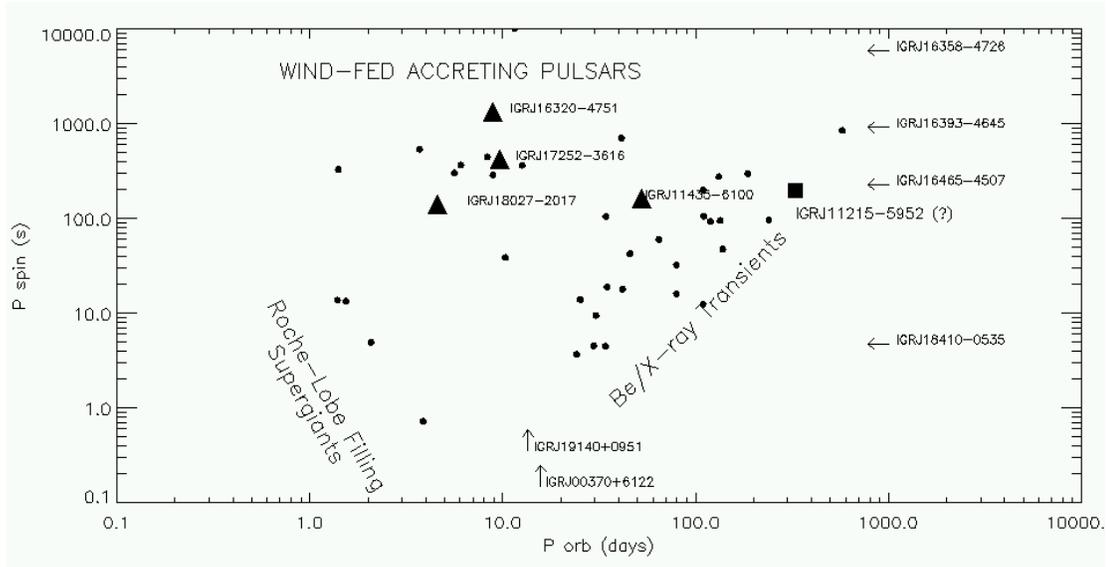}
  \caption{Corbet diagram of accreting X--ray pulsars with known orbital and spin periods.
The typical loci for wind-fed, disk-fed pulsars and transient Be/XRB are also marked. 
Few newly discovered INTEGRAL sources (IGRs) with known orbital and spin periods are also marked
with large triangles. The arrows mark the IGR sources with only one period measured (orbital or
spin period). The large square marks the  position of IGR~J11215--5952,
if the pulse period of $\sim$195~s will be confirmed.}
\end{figure}


We extracted the IBIS/ISGRI spectra at the peak of the two different
outbursts. The low energy JEM-X spectrum was available only 
for the first outburst (Figure~1). 
The 5--100~keV spectrum of July 2003 is well described 
by a
cutoff power-law ($\chi^2$=47.5 for 50
degrees of freedom, dof), with  
a photon index of 0.5$^{+0.4}_{-0.6}$ and a cutoff energy of 15$^{+5} _{-4}$~keV. 
The derived fluxes are 
6.2$\times$10$^{-10}$~erg~cm$^{-2}$~s$^{-1}$ (5--100 keV) and
2.8$\times$10$^{-10}$~erg~cm$^{-2}$~s$^{-1}$ (20--60 keV). 

During the peak of the 
second outburst (May 2004)  only the IBIS/ISGRI spectrum (20--100~keV)
was available, and could be well fitted with a single 
power-law ($\chi^2$=12.9 for 8 dof) with a photon index
of 2.6$^{+1.8} _{-0.6}$. The 20--100~keV flux is
2.5$\times$10$^{-10}$~erg~cm$^{-2}$~s$^{-1}$ and the 20--60~keV
flux is 2.1$\times$10$^{-10}$~erg~cm$^{-2}$~s$^{-1}$.
Note that the cutoff power-law best-fit to the first outburst is a good fit to the
spectrum of the second outburst as well.

\section{Discussion and Conclusions}

Analysing all the public INTEGRAL data, we have discovered
two unnoticed outbursts from the source \src. 
Our refined position was still
consistent with the proposed optical counterpart HD~306414.
Masetti et al. (\cite{Masetti2006}) found evidence for an H$_{\alpha}$ emission line, 
and confirmed  the
spectral classification as a  B1~Ia-type star, with an estimated 
distance of d$\sim$6.2~kpc.
At this distance, the peak fluxes of the two outbursts
correspond to a  luminosity of
$\sim$3$\times$10$^{36}$~erg~s$^{-1}$ (5--100~keV). This
luminosity, as well as the spectral shape derived with INTEGRAL,
are typical of  High Mass X--ray Binaries containing a neutron
star.

The three outbursts from \src\ observed with INTEGRAL 
are spaced by $\sim$330~days, possibly
indicative of an orbital period.
This suggested periodicity has been confirmed by the detection of the fourth outburst
from \src\ with RossiXTE/PCA \cite{Smith2006a} on March 16-17, 2006, 
after 329 days from the third outburst.
This observation confirmes our suggested periodicity, and makes \src\ the first SFXT
with periodic outbursts.
The X--ray emission seen with RXTE/PCA, shows a strong variability
and a hard spectrum,  well fit by a power-law with a photon index 
of 1.7$\pm{0.2}$ in the range 2.5-15 keV, absorbed by a column density of
11$\times$10$^{22}$~cm$^{-2}$, higher than the interstellar value.

Following the report of renewed activity in \src,
a Swift/XRT ToO observation of the source field was obtained \cite{Steeghs2006}, 
leading to a refined source position: 
R.A.(J2000): 11$^h$~21$^m$~46.9$^s$, Dec(J2000): --59$^{\circ}$~51$'$~42$''$ 
with an estimated error radius of 5$''$. 
This is consistent with the INTEGRAL position, and still includes the 
supposed optical counterpart, confirming the physical association with the
blue supergiant star.

The transient nature of the source, the spectral properties and
the association with a blue supergiant, confirm that \src\ is a member
of the growing class of Supergiant Fast X--ray Transients (SFXTs).
 
This is a new class of X--ray
binaries with a supergiant companion, similar to  
the persistent accreting pulsars, 
but displaying bright X--ray emission only during short X--ray outbursts. 
This transient nature is 
quite surprising since neutron stars accreting from the winds of
supergiant companions were, until recently, seen as relatively
steady sources. 
The four outbursts observed to date from \mbox{\src}
are equally spaced by about 330 days, possibly indicative of the orbital
period of the binary system.
This periodicity is worth noting, since in no
other source belonging to the class of SFXTs a periodic behavior
has been observed. 
Such a long period is more typical of Be/X--ray
binaries than of Supergiant HMXRBs, which typically have orbital
periods shorter  than $\sim$20~days. 
Smith et al. \cite{Smith2006b} reported a possible pulse period
of $\sim$195~s from \src\ detected during the RXTE/PCA observations.
If this is confirmed, \src\ would be located in the typical region 
of Be/X--ray binary pulsars in the so-called Corbet diagram
of pulse period versus orbital period for HMXRBs (see Figure~2).

The long orbital period found in \src\ is consistent with the
scenario proposed by
Negueruela et al. (\cite{Negueruela2005a}) for the SFXTs:
they suggested that SFXTs have wider orbits than ``normal'' supergiant
persistent HMXRBs (Vela X--1-like systems) and that the compact
source accretes from a less dense environment, in order to explain
the very low emission level during quiescence in SFXTs
($\sim$10$^{32}$--10$^{33}$~erg~s$^{-1}$). 
The short outbursts and the long orbital period indicate
that the binary system is wide and with a high eccentricity.


\begin{theacknowledgments}
This work has been partially supported by the Italian Space Agency
and by the MIUR under grant PRIN 2004-023189.
L. Sidoli thanks D.M. Smith for keeping informed on the source behaviour 
during the 4th outburst observed with RXTE/PCA.
\end{theacknowledgments}



\bibliographystyle{aipproc}   

\bibliography{biblio}

\begin{thebibliography}{14}
\expandafter\ifx\csname natexlab\endcsname\relax\def\natexlab#1{#1}\fi
\providecommand{\enquote}[1]{``#1''}
\expandafter\ifx\csname url\endcsname\relax
  \def\url#1{\texttt{#1}}\fi
\expandafter\ifx\csname urlprefix\endcsname\relax\def\urlprefix{URL }\fi
\providecommand{\eprint}[2][]{\url{#2}}

\bibitem[{Bird} et~al.(2006)]{Bird2006}
A.~J. {Bird}, E.~J. {Barlow}, L.~{Bassani}, A.~{Bazzano}, G.~{B{\'e}langer},
  A.~{Bodaghee}, F.~{Capitanio}, A.~J. {Dean}, M.~{Fiocchi}, A.~B. {Hill},
  F.~{Lebrun}, A.~{Malizia}, J.~M. {Mas-Hesse}, M.~{Molina}, L.~{Moran},
  M.~{Renaud}, V.~{Sguera}, S.~E. {Shaw}, J.~B. {Stephen}, R.~{Terrier},
  P.~{Ubertini}, R.~{Walter}, D.~R. {Willis}, and C.~{Winkler}, \emph{\apj}
  \textbf{636}, 765--776 (2006).

\bibitem[{Negueruela} et~al.(2006)]{Negueruela2005a}
I.~{Negueruela}, D.~M. {Smith}, P.~{Reig}, S.~{Chaty}, and J.~M.
  {Torrej{\'o}n}, \enquote{{Supergiant Fast X-ray Transients: a new class of
  high mass X-ray binaries unveiled by INTEGRAL},} in \emph{Proceedings of the
  ''The X-ray Universe 2005'', 26-30 September 2005, El Escorial, Madrid,
  Spain. Ed. by A. Wilson. ESA SP-604, Volume 1, Noordwijk: ESA Publications
  Division, ISBN 92-9092-915-4, 2006, p. 165 - 170}, edited by A.~{Wilson},
  2006, pp. 165--170.

\bibitem[{Sguera} et~al.(2005)]{Sguera2005}
V.~{Sguera}, E.~J. {Barlow}, A.~J. {Bird}, D.~J. {Clark}, A.~J. {Dean}, A.~B.
  {Hill}, L.~{Moran}, S.~E. {Shaw}, D.~R. {Willis}, A.~{Bazzano},
  P.~{Ubertini}, and A.~{Malizia}, \emph{\aap} \textbf{444}, 221--231 (2005),
  \eprint{astro-ph/0509018}.

\bibitem[{Lubinski} et~al.(2005)]{Lubinski2005}
P.~{Lubinski}, M.~G. {Bel}, A.~{von Kienlin}, C.~{Budtz-Jorgensen},
  B.~{McBreen}, P.~{Kretschmar}, W.~{Hermsen}, and P.~{Shtykovsky}, \emph{The
  Astronomer's Telegram} \textbf{469} (2005).

\bibitem[{Negueruela} et~al.(2005)]{Negueruela2005b}
I.~{Negueruela}, D.~M. {Smith}, and S.~{Chaty}, \emph{The Astronomer's
  Telegram} \textbf{470} (2005).

\bibitem[{Sidoli} et~al.(2006)]{SidoliPM2006}
L.~{Sidoli}, A.~{Paizis}, and S.~{Mereghetti}, \emph{\aap} \textbf{450},
  L9--L12 (2006), \eprint{astro-ph/0603081}.

\bibitem[{Ubertini} et~al.(2003)]{Ubertini2003}
P.~{Ubertini}, F.~{Lebrun}, G.~{Di Cocco}, A.~{Bazzano}, A.~J. {Bird},
  K.~{Broenstad}, A.~{Goldwurm}, G.~{La Rosa}, C.~{Labanti}, P.~{Laurent},
  I.~F. {Mirabel}, E.~M. {Quadrini}, B.~{Ramsey}, V.~{Reglero}, L.~{Sabau},
  B.~{Sacco}, R.~{Staubert}, L.~{Vigroux}, M.~C. {Weisskopf}, and A.~A.
  {Zdziarski}, \emph{\aap} \textbf{411}, L131--L139 (2003).

\bibitem[{Vedrenne} et~al.(2003)]{Vedrenne2003}
G.~{Vedrenne}, J.-P. {Roques}, V.~{Sch{\"o}nfelder}, P.~{Mandrou}, G.~G.
  {Lichti}, A.~{von Kienlin}, B.~{Cordier}, S.~{Schanne}, J.~{Kn{\"o}dlseder},
  G.~{Skinner}, P.~{Jean}, F.~{Sanchez}, P.~{Caraveo}, B.~{Teegarden}, P.~{von
  Ballmoos}, L.~{Bouchet}, P.~{Paul}, J.~{Matteson}, S.~{Boggs}, C.~{Wunderer},
  P.~{Leleux}, G.~{Weidenspointner}, P.~{Durouchoux}, R.~{Diehl}, A.~{Strong},
  M.~{Cass{\'e}}, M.~A. {Clair}, and Y.~{Andr{\'e}}, \emph{\aap} \textbf{411},
  L63--L70 (2003).

\bibitem[{Lund} et~al.(2003)]{Lund2003}
N.~{Lund}, C.~{Budtz-J{\o}rgensen}, N.~J. {Westergaard}, S.~{Brandt}, I.~L.
  {Rasmussen}, A.~{Hornstrup}, C.~A. {Oxborrow}, J.~{Chenevez}, P.~A. {Jensen},
  S.~{Laursen}, K.~H. {Andersen}, P.~B. {Mogensen}, I.~{Rasmussen},
  K.~{Om{\o}}, S.~M. {Pedersen}, J.~{Polny}, H.~{Andersson}, T.~{Andersson},
  V.~{K{\"a}m{\"a}r{\"a}inen}, O.~{Vilhu}, J.~{Huovelin}, S.~{Maisala},
  M.~{Morawski}, G.~{Juchnikowski}, E.~{Costa}, M.~{Feroci}, A.~{Rubini},
  M.~{Rapisarda}, E.~{Morelli}, V.~{Carassiti}, F.~{Frontera}, C.~{Pelliciari},
  G.~{Loffredo}, S.~{Mart{\'{\i}}nez N{\'u}{\~n}ez}, V.~{Reglero},
  T.~{Velasco}, S.~{Larsson}, R.~{Svensson}, A.~A. {Zdziarski},
  A.~{Castro-Tirado}, P.~{Attina}, M.~{Goria}, G.~{Giulianelli}, F.~{Cordero},
  M.~{Rezazad}, M.~{Schmidt}, R.~{Carli}, C.~{Gomez}, P.~L. {Jensen},
  G.~{Sarri}, A.~{Tiemon}, A.~{Orr}, R.~{Much}, P.~{Kretschmar}, and H.~W.
  {Schnopper}, \emph{\aap} \textbf{411}, L231--L238 (2003).

\bibitem[{Sguera} et~al.(2006)]{Sguera2006}
V.~{Sguera}, A.~{Bazzano}, A.~J. {Bird}, A.~J. {Dean}, P.~{Ubertini}, E.~J.
  {Barlow}, L.~{Bassani}, D.~J. {Clark}, A.~B. {Hill}, A.~{Malizia},
  M.~{Molina}, and J.~B. {Stephen}, \emph{\apj} \textbf{646}, 452--463 (2006),
  \eprint{astro-ph/0603756}.

\bibitem[{Masetti} et~al.(2006)]{Masetti2006}
N.~{Masetti}, M.~L. {Pretorius}, E.~{Palazzi}, L.~{Bassani}, A.~{Bazzano},
  A.~J. {Bird}, P.~A. {Charles}, A.~J. {Dean}, A.~{Malizia}, P.~{Nkundabakura},
  J.~B. {Stephen}, and P.~{Ubertini}, \emph{\aap} \textbf{449}, 1139--1149
  (2006), \eprint{astro-ph/0512399}.

\bibitem[{Smith} et~al.(2006{\natexlab{a}})]{Smith2006a}
D.~M. {Smith}, N.~{Bezayiff}, and I.~{Negueruela}, \emph{The Astronomer's
  Telegram} \textbf{766} (2006{\natexlab{a}}).

\bibitem[{Steeghs} et~al.(2006)]{Steeghs2006}
D.~{Steeghs}, M.~A.~P. {Torres}, and P.~G. {Jonker}, \emph{The Astronomer's
  Telegram} \textbf{768} (2006).

\bibitem[{Smith} et~al.(2006{\natexlab{b}})]{Smith2006b}
D.~M. {Smith}, N.~{Bezayiff}, and I.~{Negueruela}, \emph{The Astronomer's
  Telegram} \textbf{773} (2006{\natexlab{b}}).

\end{thebibliography}

\IfFileExists{\jobname.bbl}{}
 {\typeout{}
  \typeout{******************************************}
  \typeout{** Please run "bibtex \jobname" to optain}
  \typeout{** the bibliography and then re-run LaTeX}
  \typeout{** twice to fix the references!}
  \typeout{******************************************}
  \typeout{}
 }

\end{document}